# Superconductivity in a hole-doped Mott-insulating triangular adatom layer on a silicon surface

Xuefeng Wu,[1][**] Fangfei Ming,[2][**] Tyler S. Smith,[3] Guowei Liu,[1] Fei Ye,[1] Kedong Wang,[1] Steven Johnston,[3] and Hanno H. Weitering[3]

[1] *Department of Physics, Southern University of Science and Technology of China, Shenzhen, Guangdong 518055, China*

[2] *State Key Laboratory of Optoelectronic Materials and Technologies, School of Electronics and Information Technology and Guangdong Province Key Laboratory of Display Material, Sun Yat-sen University, Guangzhou 510275, China*

[3] *Department of Physics and Astronomy, The University of Tennessee, Knoxville, TN 37996, USA.*

**Adsorption of one-third monolayer of Sn on an atomically-clean Si(111) substrate produces a two-dimensional triangular adatom lattice with one unpaired electron per site. This dilute adatom reconstruction is an antiferromagnetic Mott insulator; however, the system can be modulation-doped and metallized using heavily-doped p-type Si(111) substrates. Here, we show that the hole-doped dilute adatom layer on a degenerately doped p-type Si(111) wafer is superconducting with a critical temperature of 4.7 ± 0.3 K. While a phonon-mediated coupling scenario would be consistent with the observed $T_C$, Mott correlations in the Sn-derived dangling-bond surface state could suppress the s-wave pairing channel. The latter suggests that the superconductivity in this triangular adatom lattice may be unconventional.**

The discovery of high-temperature superconductivity in doped copper-oxide (cuprate) materials [1] triggered an enormous interest in the relation between Mott physics, magnetism, and superconductivity [2]. Undoped cuprates are antiferromagnetic Mott insulators, where the Coulomb repulsion between electrons occupying the same atomic orbital prevents carriers from moving through the crystal. The introduction of electron vacancies or 'holes' into these materials, however, leads to the formation of Cooper pairs and their condensation into a macroscopically coherent superconducting quantum state. Now several decades later, there still is no consensus regarding the precise mechanism of cuprate superconductivity. Progress in this field would greatly benefit from discoveries of superconductivity in other Mott-insulating materials [3].

Here, we show that adsorption of only 1/3 monolayer of Sn atoms on a heavily boron-doped silicon (111) substrate [4] produces a strictly two-dimensional superconductor with a critical temperature $T_c$ of 4.7 ± 0.3 K that rivals that of $Na_xCoO_2 \cdot yH_2O$ [5,6]. Both systems can be viewed as close but very rare realizations of the triangular-lattice spin-1/2 Heisenberg antiferromagnet, which is a strong candidate for hosting exotic magnetism [7] and chiral superconductivity [8-11].. While the pairing symmetry in the Sn adatom layer remains to be determined, we present two possible scenarios for the observed superconductivity: interfacial electron-phonon coupling or an unconventional pairing scenario driven by Mott correlations. The latter would suggest the possibility of exploring unconventional (and possibly chiral) superconductivity [8-11], using a conventional semiconductor platform.

---

[**] These authors contributed equally.

The hole-doped Si(111)($\sqrt{3} \times \sqrt{3}$)$R30°$-Sn structure was grown on a degenerately doped p-type silicon substrate with a nominal room-temperature resistivity of 0.002 $\Omega$cm and a boron doping concentration of about $6 \times 10^{19}$ atoms/cm$^3$. The substrate was annealed to 1200 °C in ultrahigh vacuum (UHV) so as to prepare an atomically clean Si(111)($\sqrt{3} \times \sqrt{3}$)$R30°$-B surface reconstruction. Sn atoms were deposited onto the Si(111)($\sqrt{3} \times \sqrt{3}$)$R30°$-B surface from a thermal effusion cell while keeping the substrate temperature at around 600 °C. This procedure resulted in the formation of coexisting ($\sqrt{3} \times \sqrt{3}$)$R30°$-Sn and ($2\sqrt{3} \times 2\sqrt{3}$)$R30°$-Sn domains with absolute coverages of 1/3 and 1.1 ML, respectively [4,12]. The maximum domain (without internal domain boundary) size of the ($\sqrt{3} \times \sqrt{3}$)$R30°$-Sn superconducting domains is below 200×200 nm$^2$. Larger domains can be realized on lightly doped substrates. Additional details on surface preparation can be found in Ref. 4. STM data were acquired using a cryogenic STM (Unisoku) that can cool the sample and tip to 300 mK in the presence of a perpendicular magnetic field of up to 15 T. $dI/dV$ spectra were acquired using lock-in detection with a typical modulation voltage of 0.2 mV.

Figs. 1(a) and 1(b) show side and top views of the ($\sqrt{3} \times \sqrt{3}$)$R30°$-Sn surface, respectively. The Sn adatoms, indicated in orange, are located right above the atoms in the second Si layer and form an ordered ($\sqrt{3} \times \sqrt{3}$)$R30°$ superstructure relative to the hexagonal ($1 \times 1$) unit cell of the bulk-truncated substrate [13]. Each Sn adatom forms three saturated back bonds with the Si substrate atoms below, leaving one half-filled dangling bond orbital pointing towards the vacuum (Fig. 1(a)). Fig. 1(c) shows a scanning tunneling microscope (STM) image of the Sn adatoms revealing the triangular lattice symmetry [14].

The electronic structure of this interface is characterized by a half-filled dangling bond surface-state, which is located inside the band gap of the Si substrate [15]. The surface states associated with the back bonds of the Sn adatoms are fully occupied and degenerate with the bulk valence band continuum [14]. Density functional theory (DFT) calculations indicate the presence of a metallic surface-state band [16] with a Fermi surface and surface Brillouin zone as shown in Fig. 2(a).

Half-filled dangling bond states are inherently unstable, suggesting that this system may be subject to a charge ordering, a spin ordering, or a superconducting instability [17]. Indeed, the closely related 1/3 ML Sn/Ge(111) [18-20], Pb/Ge(111) [21,22], and Pb/Si(111) [23] systems all undergo a charge ordering instability, which is accompanied by a vertical rippling of the adatom layer and a tripling of the unit cell, i.e., ($\sqrt{3} \times \sqrt{3}$)$R30°$ → (3×3). Sn on Si(111) is the exception as it retains its ($\sqrt{3} \times \sqrt{3}$)$R30°$ symmetry down to $T$ = 5 K [24]. The insulating nature of this system is inconsistent with the DFT prediction [Fig. 2(a)] and has been interpreted in terms of

Mott correlations [4,16,17,24] Here, the Hubbard $U \approx 0.7$ eV while the band width of the dangling bond surface state is $W \approx 0.5$ eV. Indeed, differential conductance spectra acquired by STM reveal the presence of two Hubbard bands straddling the Fermi energy $E_F$, leaving a small band gap in between [Fig. 2(b)] [4,24]. Several photoemission studies hinted at the possibility that the ground state has row-like antiferromagnetic order [16,25,26] although there has been no direct confirmation of magnetism.

Previously, we monitored the evolution of the local density of states (LDOS) as a function of doping level by measuring the differential conductance with STM ($dI/dV \propto$ LDOS), where we introduced valence band holes using heavily boron-doped silicon substrates [4]. This method is similar to the modulation doping approach widely used in semiconductor heterostructure engineering [27]. We estimated that the highest doping level achievable with this method is about ten percent, based on the measured spectral weight transfer (see below) [4]. The $dI/dV$ spectra clearly reveal the characteristic hall marks of a doped Mott insulator, including the spectral weight transfer from the Hubbard bands to the quasi-particle states near the Fermi level (Fig. 2(c)) [4]. At the ten percent doping level, quasi-particle states acquire momentum dispersion and form a warped hexagonal Fermi contour, as expected from DFT calculations for the $(\sqrt{3} \times \sqrt{3})R30°$-Sn lattice [18,28] and observed in STM quasi-particle interference imaging [4]. However, the warping is greatly reduced in the experimental system due to the strong correlation-induced mass renormalization of the quasi-particle states [4]. In fact, the constant energy contour at 10 meV above $E_F$ is a perfect hexagon, meaning that the contour is perfectly nested [4].

An interesting feature in the data is the existence of a van Hove singularity in the tunneling spectra situated 6 meV below the Fermi level, seen as the sharp spike riding on top of the quasi particle peak (QPP) in Fig. 2(c) (Ref. 4), similar to observations for the $Ba_2Sr_2CuO_{6+x}$ cuprate superconductor [29]. This singularity originates from the flat saddle-point dispersion near the M-point in the surface state band dispersion [Fig. 2(a)]. As we will show, its presence suggests the possibility of a phonon-mediated superconducting state due to an interfacial coupling with the Si substrate; however, the strong correlations in the Sn film might suppress such a conventional mechanism. Interestingly, the geometrical frustration of the triangular lattice and strong correlations could produce a chiral $d + \mathrm{i}d$ superconducting order parameter that breaks time-reversal symmetry giving rise to interesting edge modes [8-11].

Fig. 2(d) shows the STM tunneling spectra of the $(\sqrt{3} \times \sqrt{3})R30°$-Sn system for temperatures between 0.5 and 5.5 K. The peak at about -6 meV is the van Hove singularity mentioned above. The most interesting feature, however, is the sharp tunneling dip near zero bias, which is an indicator of a possible superconducting ground state. Indeed, the gap is suppressed by the application of a perpendicular magnetic field, as shown in Fig. 2(e), and is fully quenched at a field of approximately 3 T (0.5 K; see also supplementary information [30]). These observations imply a superconducting origin of the spectroscopic gap. To obtain further insight into the superconducting state, we fit the tunneling spectra after dividing the tunneling spectra by the normal state spectrum measured at $B = 5$ T or 10 T. This procedure eliminates the complicated background of the spectrum created by the nearby van Hove singularity. The result, shown in Fig. 3(a), is a series of characteristic superconducting tunneling spectra. We fit the data using the Dynes formula [31] assuming either $s$- and $d+id$ wave superconducting order parameters (supplementary information [30]; see also Ref. 32). Both choices give comparable fits due to the large values of the broadening parameter $\Gamma$. The corresponding gap values $\Delta(T)$ are plotted in Fig. 3(b) and fit with the BCS mean-field temperature-dependence to yield a superconducting critical temperature $T_c = 4.7 \pm 0.3$ K. A very similar value was obtained by plotting the zero-bias conductance (ZBC) as a function of temperature [30]. Note that the data points at 5.0 and 5.5 K suggest the presence of a very small gap above 4.7 K, albeit with a large error bar. This suggests that $\Delta(T)$ deviates from conventional BCS behavior and/or that the superconducting transition is broadened due to the finite size of the superconducting domains [33].

While Fig. 1c clearly shows an undistorted $(\sqrt{3}\times\sqrt{3})R30°$-Sn lattice, to definitively rule out the possibility of a spin-wave or charge-ordering scenario, we recorded $dI/dV$ maps at 1.0 T [Fig. 4(b)] which reveal the presence of superconducting vortices. These are the yellow circular regions in Fig. 4(b), which represent regions with high ZBC. The blue regions represent the superconducting state with low ZBC. Inside each vortex, the superconducting order parameter should be suppressed. We recorded a series of tunneling spectra as a function of radial distance from the vortex center, showing the zero bias dip feature is reduced at the vortex center. We also plotted the line profiles from the $dI/dV|_{V=0}$ map in Fig. 4(b), as shown in Fig. 4(d). Indeed, the ZBC reaches a maximum at $r = 0$ and decays monotonically away from the center. We fit the resulting data using $ZBC(r) = ZBC(\infty) + Ae^{-|r-c|/\xi}$ (Ref. 34, where $c$ is the center position) to estimate a superconducting coherence length $\xi \approx 14.3 \pm 2.0$ nm at 0.5 K. The rather large value of the coherence length is consistent with the low $T_c$. A somewhat smaller value of about 10.5 nm is obtained using our upper critical field estimate $H_{c2}(0.5\text{ K}) \approx 3$ T and the Ginzburg-Landau expression $H_{c2}(T) = \Phi_0/2\pi\xi^2(T)$, where $\Phi_0 = 2.07\times 10^{-15}$ Wb is the flux quantum [35].

Superconductivity in monatomic adatom layers on Si(111) has been observed before, but only in dense monolayer systems with $T_c$ values well below the bulk $T_c$ of the corresponding adsorbate species [36-39]. Here, we have a unique situation where only 1/3 ML of Sn induces superconductivity with a $T_c$ that is comparable to and even slightly exceeding that of bulk (white) Sn ($T_c = 3.7$ K). While degenerate levels of boron in Si can produce superconductivity, the corresponding $T_c$ is only 0.35 K [40]. Indeed, we found no superconductivity in boron-doped Si(111) and no proximity-induced superconductivity in the neighboring $(2\sqrt{3}\times 2\sqrt{3})$ domains [30]. This result conclusively demonstrates that the pairing interactions only involve electrons in the dangling-bond surface state of the $(\sqrt{3}\times\sqrt{3})$-Sn structure.

The van Hove singularity at 6 meV below the Fermi level [4] is well within the phonon band width of the $(\sqrt{3}\times\sqrt{3})R30°$-Sn structure [41-43]. Using the band dispersion from Ref. 16 and published values of the electron-phonon coupling constants for this surface [43], we estimate the phonon-mediated transition temperature to be of the order of 10 K, in good agreement with the experimental result [30]. In this case, the dimensionless electron-phonon coupling constant $\lambda \cong$ 1.3 of the strongly coupled 'wagging mode' [43] is rather large, which can be attributed to the close proximity of the van Hove singularity to the Fermi level. Nonetheless, the strong Hubbard correlations and long-range Coulomb interactions could suppress the conventional s-wave pairing channel, and instead favor an unconventional $d + id$ pairing symmetry [10]. Since both order parameters were consistent with our data [Fig. 3(c)], the pairing symmetry remains to be established.

One of the biggest surprises of the hole-doped Si(111)($\sqrt{3} \times \sqrt{3}$)-Sn system is that it reproduces much of the correlated electron physics seen in complex oxides. In fact, its $T_c$ is comparable to that of the few other known triangular-lattice superconductors, including the layered hydrated sodium cobaltate $Na_xCoO_2 \cdot yH_2O$ ($T_c = 4.3$ K) [5,6] and the layered organic charge-transfer salts with κ-(ET)$_2$Cu[N(CN)$_2$]Br holding the record $T_c = 11.6$ K (Ref. 44). In particular, the sodium cobaltate compound is viewed as the triangular-lattice variant of the high-$T_c$ square-lattice cuprate superconductors [5,6]. Yet the Si(111)($\sqrt{3} \times \sqrt{3}$)-Sn system represents a class of materials that is much simpler, both chemically and electronically, and is furthermore compatible with conventional semiconductor platforms. Unlike the other systems mentioned above, it is a strictly two-dimensional single-band Mott insulator and there are no complicating factors related to structural and electronic inhomogeneities. If an unconventional pairing symmetry can be established for this system, it would provide the cleanest platform for studying superconductivity arising from a doped Mott insulator [2].

A most intriguing question is if it would be possible to enhance $T_c$ using higher hole doping levels. Recent theoretical studies of the ($\sqrt{3} \times \sqrt{3}$)-Sn system suggested the presence of a high-$T_c$ superconducting dome at doping levels above twenty percent [10]. Ten percent was the highest doping level we could achieve using commercially available heavily boron-doped substrates. It is quite conceivable that higher doping levels can be achieved using special made substrates with very heavy ion implantation. Attempts to electron-dope this system via low-dose potassium deposition resulted in local charge ordering transition near isolated adsorbate atoms [45]. Increased deposition amounts produced long-range charge order with three distinct sublattices: triangular, honeycomb, and Kagome [45]. Other Mott insulating semiconductor surfaces, such as Sn on Ge(111) [18,19], SiC(0001) [46], Sn (and possibly Pb) on SiC(0001) [47] may also become superconducting with modulation hole doping. Alternatively, one could imagine tuning the energy balance between the Mott phases and charge ordered phases in Sn/Ge(111) [20], Pb/Ge(111) [21] or Pb/Si(111) [23,48] via strain engineering using, e.g., Si/Ge alloy substrates. Dopants can be introduced via modulation doping [4,27] or deposition of electronegative (-positive) atomic or molecular species. Much work remains to be done to map the electronic phase diagrams of these triangular lattice systems, both as a function of doping and strain, and to fully elucidate the mechanism of superconductivity in these systems.

The authors acknowledge P.C. Snijders for his scientific contributions during the initial stages of the project. This work was supported by the National Science Foundation under Grant No. DMR 1410265 (H. H. W.), the Office of Naval Research under Grant No. N00014-18-1-2675 (S. J.), the National Natural Science Foundation of China Grant No. 11574128, the MOST 973 Program Grants No. 2014CB921402 (K. W.), and the Hundred Talents Plan of Sun Yat-Sen University Grant No. 76120-18841210 (F. M.). T. S. S. acknowledges additional support from the Center for Materials Processing, a Tennessee Higher Education Commission supported Accomplished Center of Excellence.

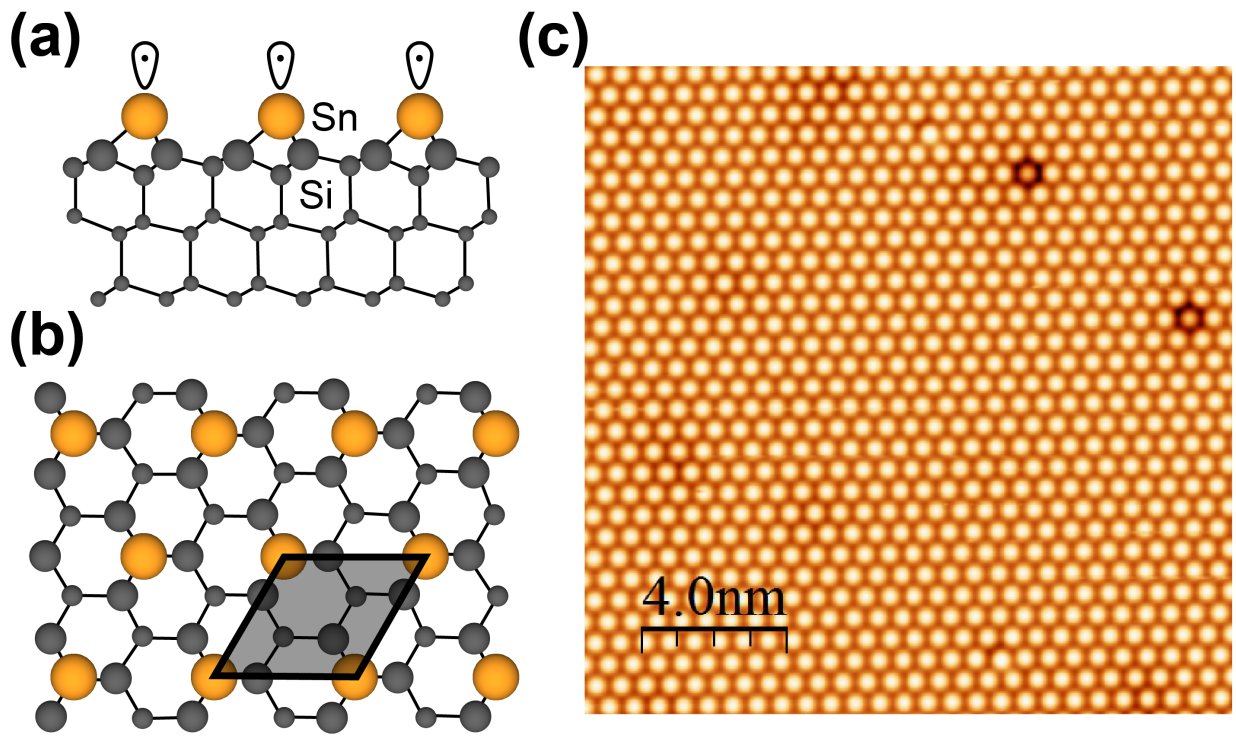

FIG. 1 Structure and STM topographic image of the Si(111)($\sqrt{3} \times \sqrt{3}$)-Sn surface reconstruction. (a) Side view of the Si(111)($\sqrt{3} \times \sqrt{3}$)-Sn surface. The Sn atoms (orange) are adsorbed at the $T_4$ adsorption sites, above the atoms of the second Si layer. Each Sn atom has a dangling bond pointing to the vacuum side containing one electron, as indicated by the small dots. (b) Top view of the Si(111)($\sqrt{3} \times \sqrt{3}$)-Sn surface. The Sn atoms form a $(\sqrt{3} \times \sqrt{3})R30°$ superstructure relative to the (1×1) periodicity of the Si(111) surface. The hexagonal $(\sqrt{3} \times \sqrt{3})R30°$ supercell is indicated by the shaded diamond. **c,** Topographic STM image of the Si(111)($\sqrt{3} \times \sqrt{3}$)-Sn surface, acquired at a tunneling bias of 1 V and a tunneling current of 1 nA. Only the Sn atoms can be seen. The two darker point-defects in the upper right corner of the image are substitutional Si atoms.

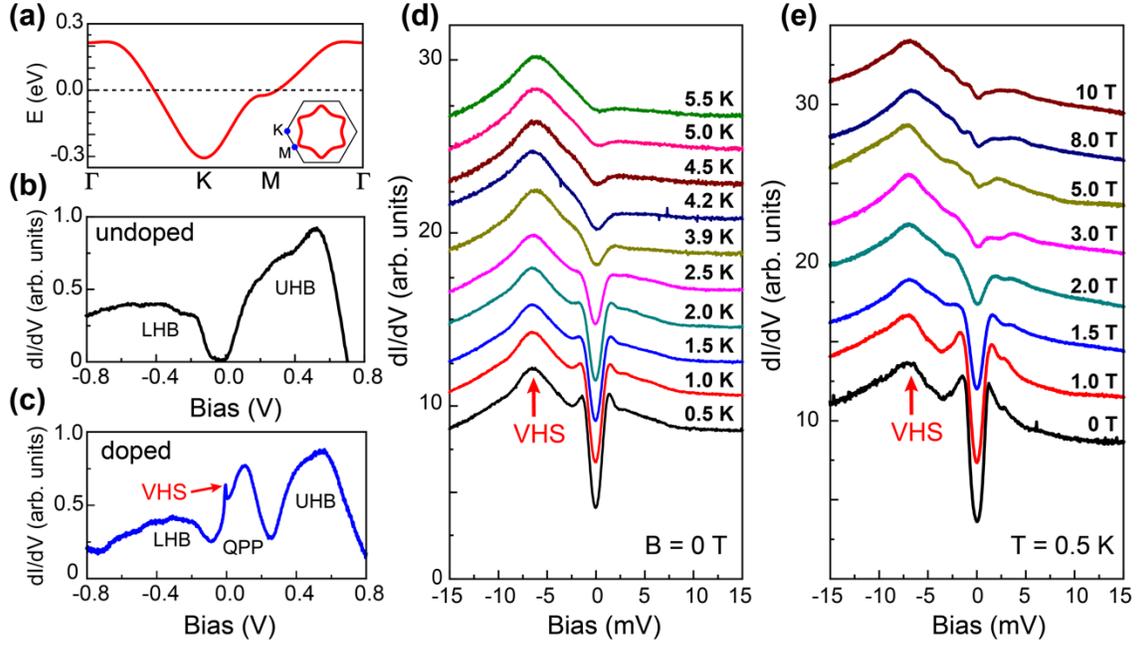

FIG. 2 Electronic structure and tunneling spectra of the Si(111)($\sqrt{3} \times \sqrt{3}$)-Sn surface. (a) Momentum dispersion of the dangling-bond surface state according to Ref. 16. Note the saddle point near the M-point of the surface Brillouin zone, which is located just below the Fermi level. The inset shows the Fermi contour (red) along with the hexagonal surface Brillouin zone. The high symmetry points are indicated. The Γ point is located at the center of the hexagon. (b) Differential conductance spectra ($dI/dV \propto$ LDOS) of a minimally doped Si(111)($\sqrt{3} \times \sqrt{3}$)-Sn surface, acquired at 77 K, showing the upper and lower Hubbard bands (UHB/LHB) and a small gap around the Fermi level [4]. (c) Same as (b) but with a hole-doping level of about ten percent [4] and measured at 5 K. Hole doping results in a transfer of spectral weight from the Hubbard bands to a new quasi particle peak (QPP) near the Fermi level. The sharp spike just below the Fermi level is the van Hove singularity (VHS). (d) $dI/dV$ spectra as a function of temperature. The peak at -6 meV is the van Hove singularity and the dip at zero bias is the superconducting gap. (e) $dI/dV$ spectra measured at 0.5 K in different perpendicular magnetic fields. The spectra in (d) and (e) are shifted vertically for clarity.

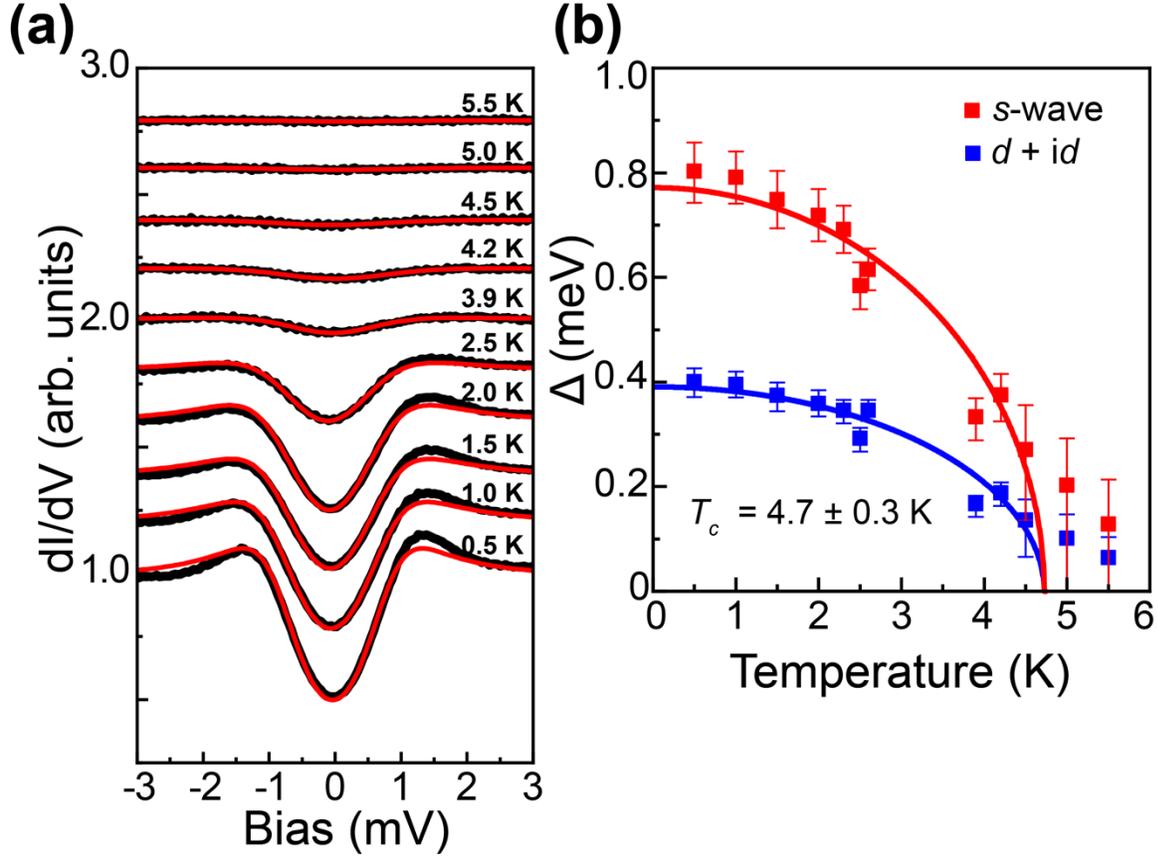

FIG. 3 Temperature-dependence of the superconducting gap. (a) Normalized $dI/dV$ spectra obtained by dividing the zero-field spectra in Fig. 2(d) by the corresponding spectra measured in a 5 T or 10 T field [30] The spectra are fitted using the Dynes formula [31] for an s-wave order parameter. Fits for the $d + \mathrm{i}d$ order parameter are shown in the supplementary information [30]. (b) $\Delta(T)$ data points obtained from the s-wave and chiral d-wave fits of the normalized $dI/dV$ spectra in (a). The solid lines are the corresponding data fits according to the BCS mean field formula $\Delta(T) = \Delta_0 \tanh\left(A\sqrt{1-(T/T_c)^2}\right)$, where $\Delta_0$, $A$, and $T_c$ are fitting parameters. Note that the data points at 5.0 K and 5.5 K seem to deviate from the BCS curve, indicating that the superconductivity is either non-BCS like and/or that there are superconducting fluctuations above the critical temperature $T_c = 4.7 \pm 0.3$ K obtained from the BCS fit.

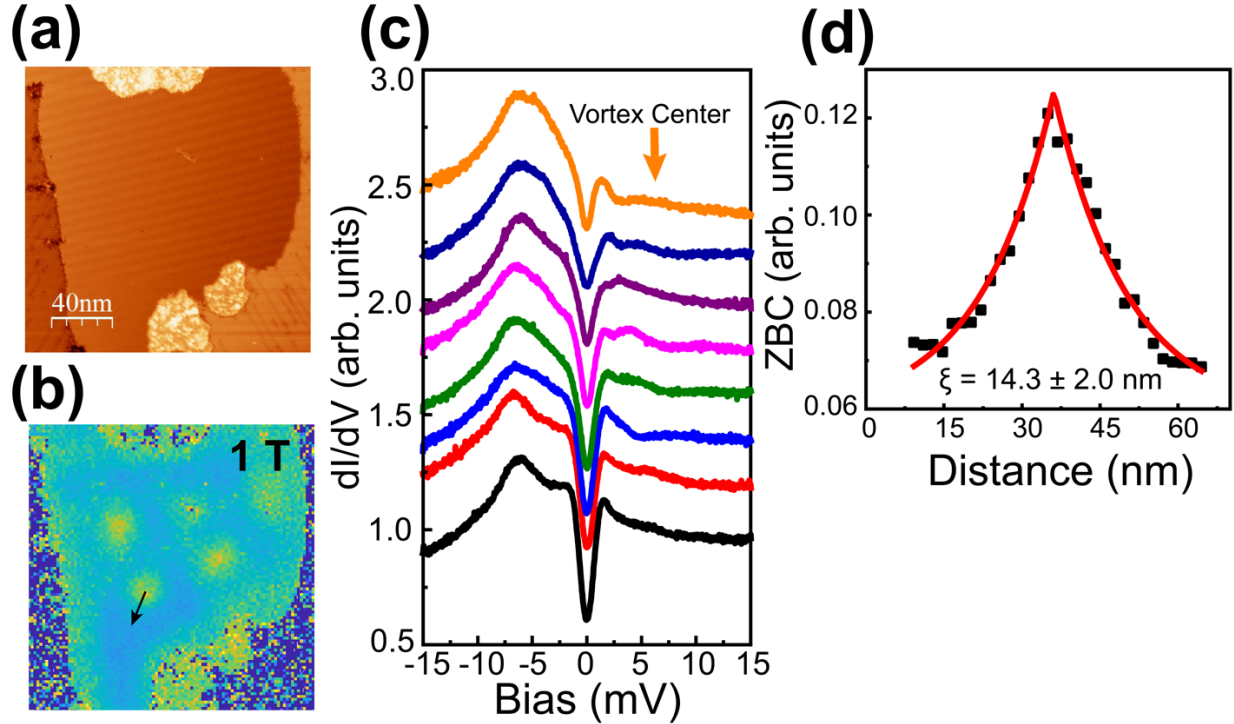

FIG. 4 Magnetic vortices and superconducting coherence length. (a) Topographic STM image of a Si(111)($\sqrt{3} \times \sqrt{3}$)-Sn surface domain, surrounded by other semiconducting phases. (b) Corresponding $dI/dV$ map, recorded at zero bias and 0.5 K in 1 T magnetic field. Regions with high (low) zero-bias conductance are indicated in yellow (blue). The yellow round patches are normal state regions due to the local penetration of magnetic flux quanta. (c) $dI/dV$ spectra recorded from along the radial direction of the vortex region, as indicated by the arrow in panel (b). Note the gradual decrease in ZBC towards the edge. (d) Averaged line profile across the vortices in (a). The exponential fit produces a coherence length $\xi \approx 14.3 \pm 2.0$ nm (0.5 K).